\documentclass{article}
\usepackage{spconf,amsmath,graphicx,bm}
\usepackage{multirow}

\title{Utterance-level end-to-end language identification using attention-based CNN-BLSTM}
\name{Weicheng Cai$^{1,2}$,Danwei Cai$^{1}$,~Shen Huang$^{3}$~and Ming Li$^{1}$\sthanks{This research was funded in part by the National Natural Science Foundation of China (61773413), Natural Science Foundation of Guangzhou City (201707010363), and CCF-Tencent Open Fund.}}

\address{$^1$Data Science Research Center, Duke Kunshan University, Kunshan, China\\
    $^2$School of Electronics and Information Technology, Sun Yat-sen University, Guangzhou, China\\
    $^3$Tencent Research, Beijing, China\\
    {\small \tt ml442@duke.edu}}

\begin{document}
\ninept
\maketitle
\begin{abstract}
In this paper, we present an end-to-end language identification framework, the attention-based~Convolutional Neural Network-Bidirectional Long-short Term Memory~(CNN-BLSTM). The model is performed on the utterance level, which means the utterance-level decision can be directly obtained from the output of the neural network. To handle speech utterances with entire arbitrary and potentially long duration, we combine~CNN-BLSTM model with a self-attentive pooling layer together.  The front-end CNN-BLSTM  module plays a role as local pattern extractor for the variable-length inputs, and the following self-attentive pooling layer is built on top to get the fixed-dimensional utterance-level representation. We conducted experiments on
NIST~LRE07 closed-set task, 
and the results reveal that the proposed attention-based CNN-BLSTM model achieves comparable error reduction
with other state-of-the-art utterance-level neural network approaches for all 3 seconds,~10 seconds,~ 30 seconds duration tasks.

\end{abstract}

\section{Introduction}
Language identification~(LID) can be defined as an utterance-level ``variable-length sequence classification" task. It is a problem in that we are trying to retrieve information about an entire utterance rather than specific word content~\cite{campbell2006support}. Moreover, there is no constraint on the lexicon words thus training utterances and testing segments may have completely different contents~\cite{Kinnunen2010An}. Therefore, given the input speech utterances of arbitrary duration, our goal may boil down to transform them into fixed-dimensional representations, among them the inter-class variability is maximized and simultaneously the intra-class variability is minimized~\cite{hansen2015speaker}.

There are generally two categories to obtain the fixed-dimensional utterance-level representations. The first comprises stacking self-contained algorithmic components. The representative is the classical i-vector approach~\cite{dehak2010front}. Firstly, variable-length feature sequences are extracted from raw audio signals. Then, selected feature frames in the training dataset are grouped together to estimate a Gaussian Mixture Model~(GMM) based universal background model~(UBM)~\cite{Reynolds2000Speaker}. Sufficient statistics of each utterance on the UBM is accumulated, and a factor analysis based i-vector extractor is trained to project the statistics into a low rank total variability subspace.  The main advantage of this type of methods is that the system can accept variable-length input. Any input speech segment of arbitrary duration can be transformed as a fixed-dimensional i-vector representation.

Another category relies on the model trained by a downstream procedure through a deep neural network (DNN). In the early stages, the DNN-based LID model commonly performs prediction at the frame level, and the original input feature sequences are resized or cropped into multiple small ﬁxed-length segments, as is done in~\cite{lopez2014automatic,gonzalez2014automatic, Li2016Exploiting, 10.1007/978-3-319-43958-7_53}. Since the DNNs only provide frame-level prediction., the final utterance level scores are derived by averaging the frame-level posteriors.

To handle the variable-length speech utterances,  recently, several context-independent pooling layers such as temporal average pooling (TAP)~\cite{caie2e_iccasp18,snyder2018spoken} layer, self-attentive pooling~(SAP) layer~\cite{Cai_2018_Odyssey,is18_attentive}, and learnable dictionary encoding~(LDE)~\cite{cailde_iccasp18} layer has been introduced for the utterance-level LID modeling. The pooling layer is first introduced to speaker recognition task~\cite{1705.02304, Snyder2017Deep}, and typically built on top of the front-end CNN or time-delay neural network~(TDNN)~\cite{waibel1990phoneme} to get the utterance-level representation within the network structure. With the merit of pooling layer, the DNNs can train input segments with variable duration. In the testing stage, the whole speech utterances of arbitrary duration can be fed into the DNNs directly. 

\begin{figure*}[tb]
        \centering    
        \includegraphics[width=0.91\textwidth]{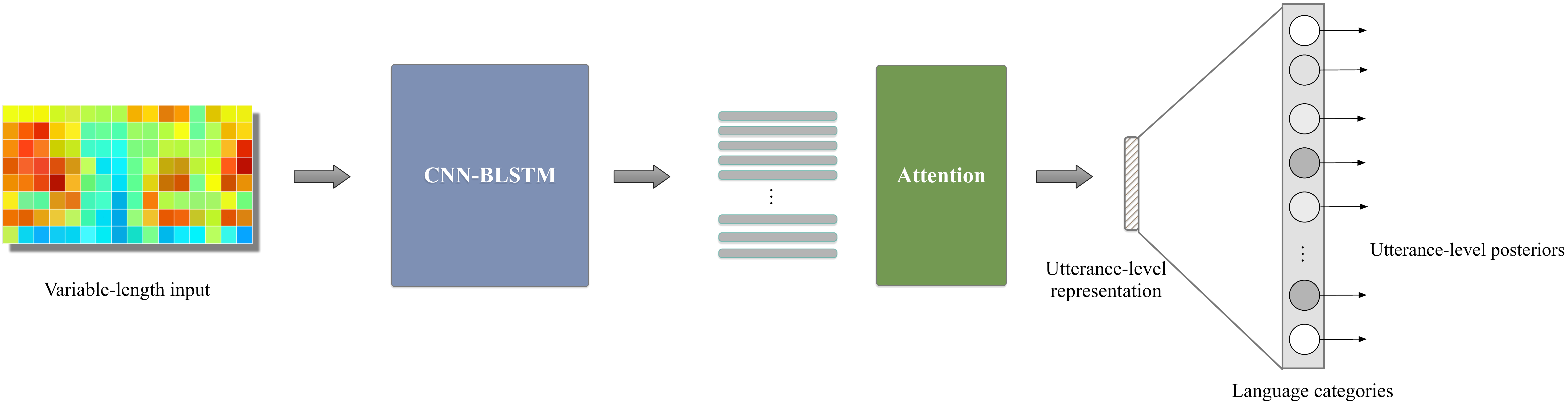}
        \caption{Our proposed attention-based CNN-BLSTM framework for LID.  It accepts input data sequence with variable length, and  produces an utterance-level result directly from the output of the DNN.}\label{fig:system}
 \end{figure*}

 \begin{figure}[tb]
        \centering    
        \includegraphics[width=0.58\columnwidth]{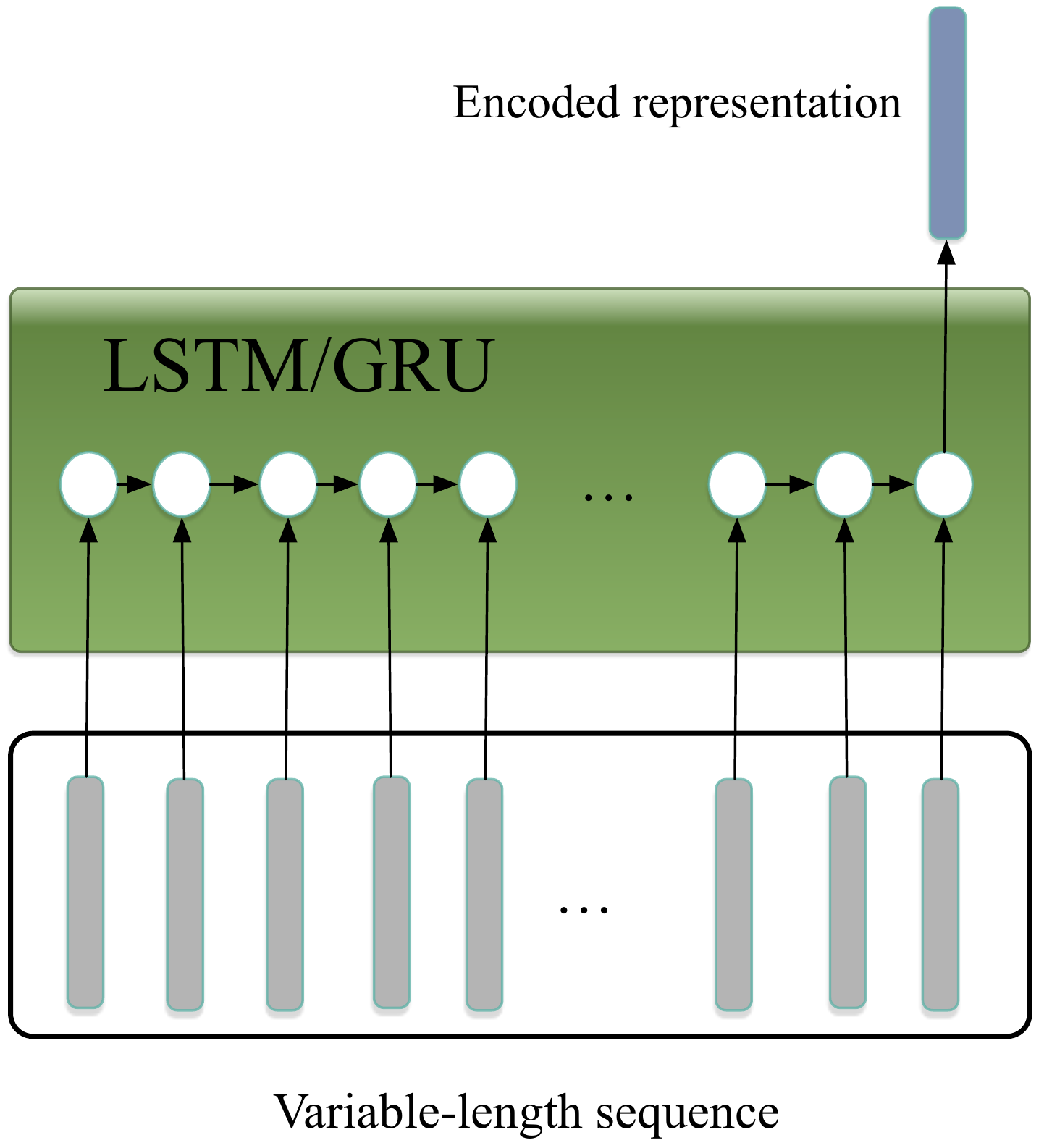}
        \caption{Diagram of the recurrent encoding layer}\label{fig:recurrent}
\end{figure}
  
Besides the context-independent pooling layer, theoretically, we can also get the fixed-dimensional utterance-level representation through context-dependent recurrent manner. Since recurrent layer can make full use of the context of feature sequences in the forward direction,  in our previous work~\cite{caie2e_iccasp18}, we applied recurrent layer such as LSTM or Gated Recurrent Unit (GRU) to get the encoded utterance-level representation. We regard the last output vector of the LSTM/GRU layer as the encoded representation and put it into the subsequent fully-connected~(FC) layer, as demonstrated in Fig.~\ref{fig:recurrent}.  It is surprised that although LSTM or GRU layers introduce much more parameter than TAP layer, it results in degraded performance, especially for the testing task over a long-range duration~\cite{caie2e_iccasp18}.

The recurrent layer may suffer from ``the curse of sentence length''~\cite{cho2014properties} when it plays a role as encoding layer. However, we could not neglect the power of the recurrent layer in modeling the temporal dynamic behavior for the time sequence. This kind of context-dependent temporal structure may be useful for recognizing spoken language. Therefore, motivated by the success of connecting CNN and recurrent layer in some speech recognition~\cite{sainath2015convolutional,xiong2018microsoft} and neural language understanding~\cite{shen2017sentiment} tasks, we replace the front-end CNN module with a tandem CNN-BLSTM structure. Rather than directly using the output of the last time step as the encoded output information, we employ the recurrent layer as part of our front-end local pattern extractor. Based on the tandem CNN-BLSTM,  we additionally apply another attention-based pooling layer on the outputs of all time steps to extract the fixed-dimensional representation. The model is performed on the utterance level, which means the utterance-level decision can be directly obtained from the output of the neural network, given variable-length input sequences.

\section{Attention-based CNN-BLSTM}
\label{sec:overview}
    
\subsection{System overview}

The speech signal is naturally with variable length, and we usually don't know exactly how long the testing speech segment will be. Therefore, a flexible processing method might be able to accept speech segments of arbitrary duration. Fig. \ref{fig:system} shows our utterance-level DNN architecture. 

The front-end tandem CNN-BLSTM module plays a role as local pattern extractor for the variable-length input sequence.   The CNN-BLSTM learned high-level abstract pattern is still with temporal order. The remaining question is: how to aggregate them together over the entire and potentially long duration?  Concerning about that, an SAP layer is designated on top of the CNN-BLSTM to get the fixed-dimensional utterance-level representation. The utterance-level representation after the SAP layer can be further processed through an FC layer and finally connected with a classification output layer. Each unit in the output layer is represented as a target language category. 

 After model have been trained, given testing utterances of arbitrary duration, the utterance-level posteriors could be directly obtained from the output of the neural network.

\subsection{Tandem CNN-BLSTM}    
\subsubsection{CNN}

\begin{figure}[tb]
        \centering    
        \includegraphics[width=0.92\columnwidth]{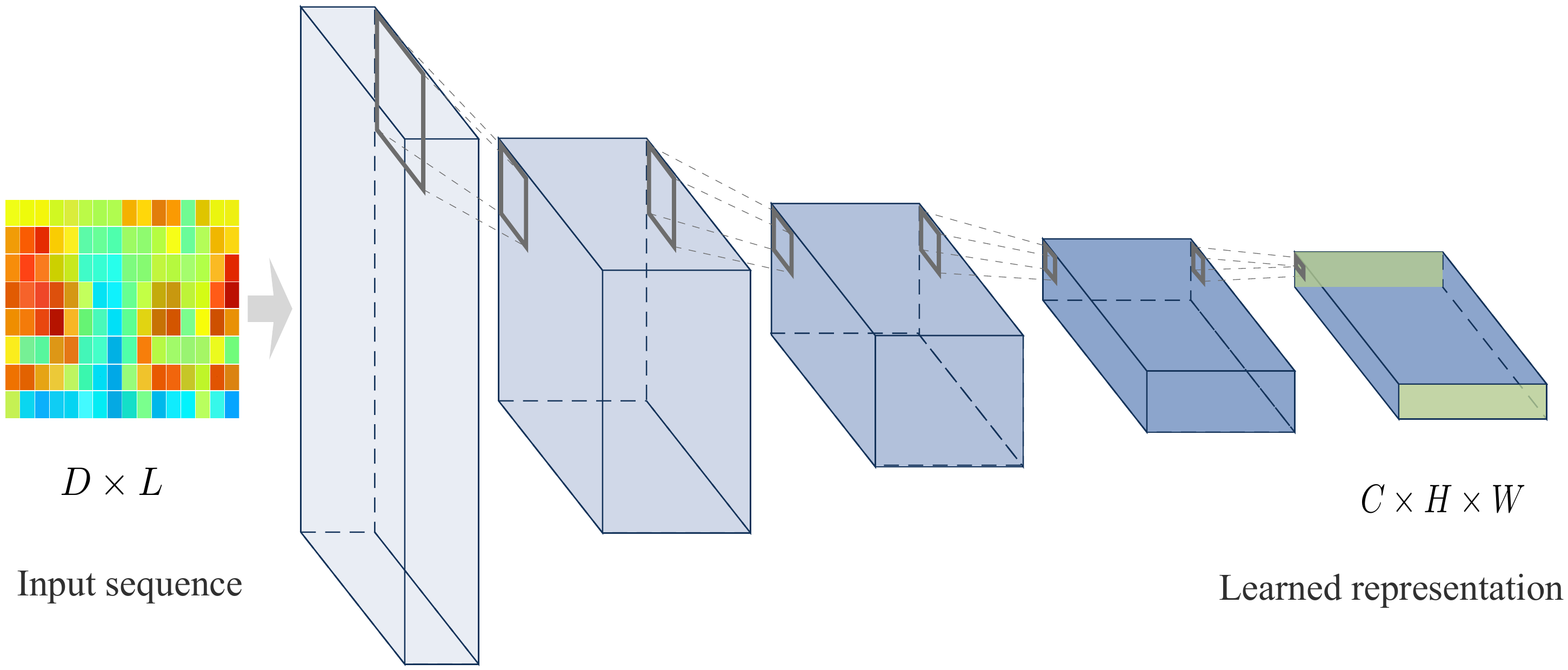}
        \caption{The input-out structure of CNN. It receives a two-dimensional feature matrix $D\times L$, and produces a 3D tensor block $C\times H \times W$}\label{fig:cnn}
\end{figure}
CNN is a kind of specialized neural network for processing data with a known grid-like topology, and it uses learned filters to convolve the feature maps from the previous layer. To some extent, therefore, the convolution layer of CNNs operates in a sliding window manner acting as an automatic local feature extractor. 

For a given input temporal ordered feature sequence with the shape $D \times L$ (where $D$ denotes the feature dimension of the acoustic features, and $L$ denotes the number of frames), typically, as described in Fig. \ref{fig:cnn}， the CNN learned representations are a three-dimensional tensor block with the shape of $C\times W\times H$, where $C$ denotes the number of channels, $H$ and $W$ denotes the height and width of the feature maps. Generally, $W$ and $H$ are much smaller than the original $D$ and $L$, since we have many downsample operations within the CNN structure.

\subsubsection{BLSTM}

For LID task, it might be beneficial
to have access to future as well as past context.
However, standard LSTM networks ignore future context.
BLSTM extend the unidirectional
LSTM networks by introducing a backward direction layer. The model is, therefore, able to exploit information both from the
past and future \cite{schuster1997bidirectional}.

    The CNNs produce the three-dimensional output with the shape of $C\times H \times W$. However,  BLSTM expects inputs to be a two-dimensional tensor. Therefore, we first pool the output of CNN along its height axis and then squeeze it into a two-dimensional $C \times W$ representation by removing the dimension of size 1. Here, $C$ not only denotes the number of channels of CNN but also represents the input feature dimension for BLSTM. Similarly,  $W$ not only denotes the width of the CNN feature maps but also represents the time step in BLSTM. It is noticed that $W$ is a variable number considering different input frames $L$. 

\begin{figure}[tb]
        \centering    
        \includegraphics[width=0.5\columnwidth]{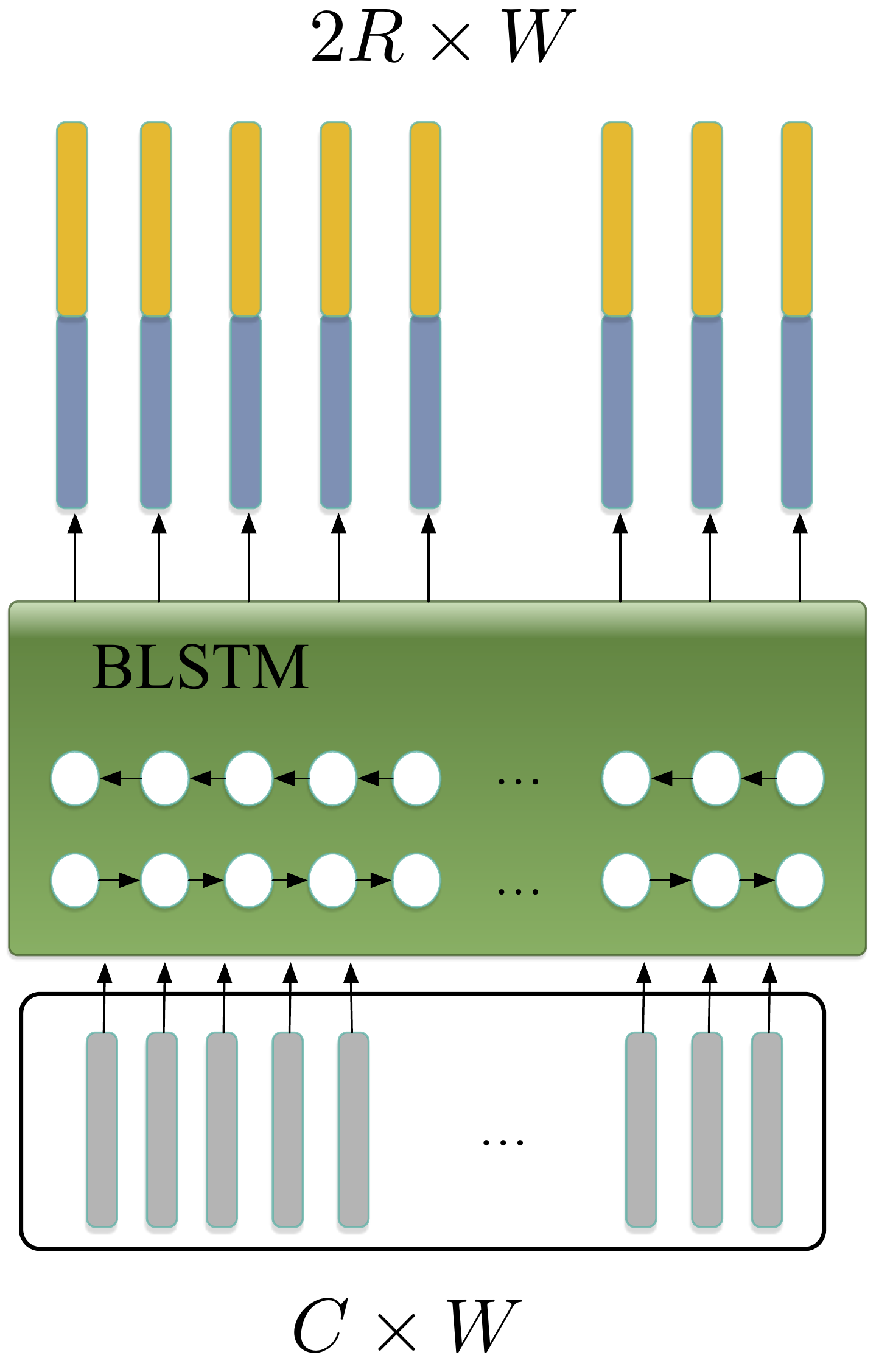}
        \caption{The input-out structure of BLSTM layer. It receives a variable-length sequence $C\times W$, also produces a variable-length sequence $2R \times W$}\label{fig:blstm}
\end{figure}

As shown in Fig. \ref{fig:blstm}, the output of the BLSTM is also a variable-length sequence of shape $2R \times W$, where $R$ denotes the output feature dimension of the BLSTM. 

\subsection{Attention mechanism}
\begin{figure}[tb]
        \centering    
        \includegraphics[width=0.63\columnwidth]{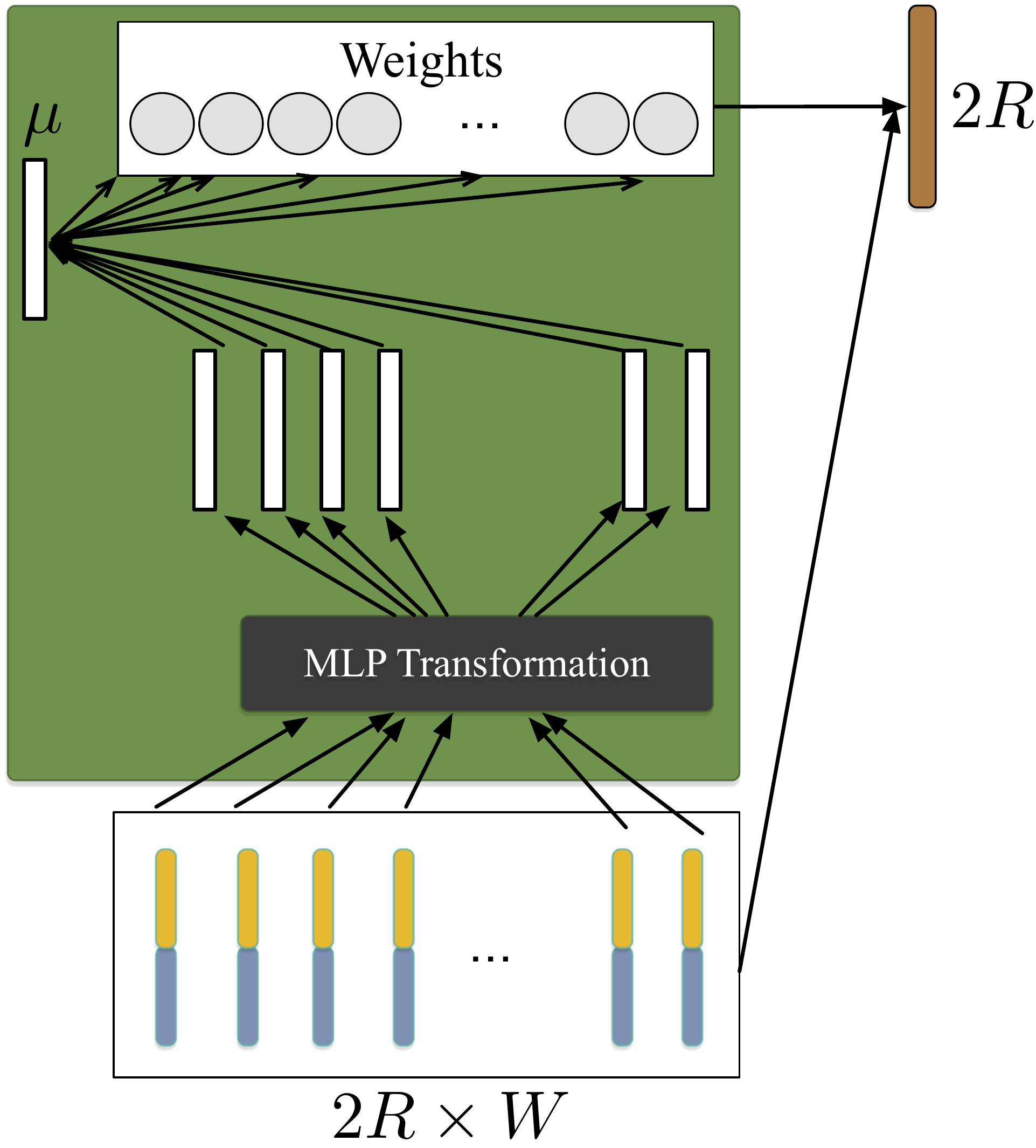}
        \caption{The input-out structure of SAP layer. It receives a variable-length sequence $2R \times W$, produces a fixed-dimensional utterance-level representation of size $2R$}\label{fig:SAP}
\end{figure}

In our previous work, we employ a TAP layer on top of the CNN and pool the CNN extracted feature maps equally.  Considering the BLSTM produced variable-length sequences, we can also use the same TAP layer to aggregate the variable-length sequence together. However, not all frame of features contributes equally to the utterance-level representation. For example,  the last time step in the forward direction and the first step in backward direction may be encoded more information about the sequence. Therefore, we introduce an attention-based pooling layer to pay attention to such frames that are important to the classification and aggregate those informative frames to form an utterance-level representation.  

We implement the SAP layer the same as in~\cite{Cai_2018_Odyssey}. That is, we first feed the variable-length sequence produced by BLSTM $\{\bm{x_{1}},\bm{x_{2}},\cdots,\bm{x_{T}}\}$  into 
a multi-layer  perceptron (MLP)  to get $\{\bm{h_{1}},\bm{h_{2}},\cdots,\bm{h_{T}}\}$ as a hidden representation. In this paper, we simply adopt  a  one-layer perceptron, 
\begin{equation}
\bm{h}_{t} = \tanh(\bm{\mathcal{W}x}_{t} + \bm{b})
\end{equation}
Then we measure the importance of each frame as the similarity of $\bm{h_{t}}$ with a learnable 
context vector $\bm{\mu}$ and get a normalized importance
weight $\alpha_t$ through a softmax function.
\begin{equation}
\alpha_{t} = \frac{\exp(\bm{h}_{t}^T\bm{\mu})}{\sum_{t=1}^{T}\exp({\bm{h}_{t}^T}\bm{\mu})}
\end{equation}

The context vector $\bm{\mu}$
can be seen as a high-level representation of a fixed
query ``what is the informative frame” over the whole frames~\cite{yang2016hierarchical}. It is randomly initialized and jointly learned
during the training process.

After that,  as demonstrated in Fig.~\ref{fig:SAP}, the utterance-level representation $\bm{e}$  can be generated as a weighted sum of the BLSTM produced variable-length sequence 
based on the learned weights $\alpha_t$, 
\begin{equation}
\bm{e} = \sum_{t=1}^{T}\alpha_t \bm{x}_{t}
\end{equation}

\begin{table*} [tb]
    \caption{  Performance on the 2007 NIST LRE closed-set task. N/R: Not reported}
    \centerline{
        \begin{tabular}{c c c c c c c c c c c }
            \hline
            { System}& \multirow{2}{*}{ System Description}&\multirow{2}{*}{ Front-end module}&\multirow{2}{*}{Encoding layer}&\multicolumn{3}{c}{$C_{avg}(\%)$}&\multicolumn{3}{c}{$EER(\%)$}\\
            \cline{5-10}
            ID&&&&3s&10s&30s&3s&10s&30s\\
            \hline
            1& CNN-TAP~\cite{caie2e_iccasp18}  &CNN&TAP&9.98&3.24&1.73&11.28&5.76&3.96\\    
            2& CNN-SAP~\cite{Cai_2018_Odyssey}  &CNN&SAP&\textbf{8.59}&\textbf{2.49}&1.09&9,89&4.27&2.38\\
            3& CNN-LSTM~\cite{caie2e_iccasp18}  &CNN&LSTM&10.17&4.66& N/R &9.80&4.26&N/R\\    
            4& CNN-GRU~\cite{caie2e_iccasp18}  &CNN&GRU&11.31&5.49&N/R&10.74&6.40&N/R\\        
                5& LSTM-Attention~\cite{Geng2016End}  &LSTM&Attention&N/R&N/R&N/R&14.72&N/R&N/R\\        
            6& CNN-BLSTM TAP  &CNN-BLSTM&TAP&9.83&3.31&2.03&11.22&5.26&3.67\\    
            7& \textbf{CNN-BLSTM SAP} &CNN-BLSTM & SAP& 9.22&2.54&\textbf{0.97}&\textbf{9.50}&\textbf{3.48}&\textbf{1.77}\\
    \hline
            8& \textbf{Fusion ID2 + ID7} &&&\textbf{7.98} &\textbf{2.30}& \textbf{0.89}&\textbf{8.03}&\textbf{3.05}&\textbf{1.56}\\
            \hline
    \end{tabular}}
    \label{table:lre07}
\end{table*}

\begin{table}[tb]
    \normalsize
    \centering
    \caption{Attention-based CNN-BLSTM network structure. N/A: Not available}
    \label{resnetconfig}
    \resizebox{0.98\columnwidth}{!}{
        \renewcommand\arraystretch{1.3}
        \begin{tabular}{|c|c|c|c|c|}
            \hline
            Layer               & Output size            & Downsample      &  Channels     &  Blocks      \\ \hline
            Conv1                     & 64 $\times$ $L$                & False& 16       & -    \\ \hline
            Res1                & 64$\times$  $L$               & False& 16& 3    \\ \hline
            Res2                & 32 $\times$ $\frac{L}{2}$               & True& 32& 4    \\ \hline
            Res3                & 16 $\times$ $\frac{L}{4}$              & True& 64& 6   \\ \hline
            Res4                & 8 $\times$ $\frac{L}{8} $              & True& 128& 3    \\ \hline
            Pool         & 128$\times$ $\frac{L}{8} $  &N/A&N/A&N/A \\ \hline        
                BLSTM         & 256$\times$ $\frac{L}{8} $  &N/A&N/A&N/A \\ \hline    
            SAP        & 256 &N/A&N/A&N/A \\ \hline    
            Output         &  14 &N/A&N/A&N/A \\ \hline    
    \end{tabular}}
\end{table}

    \section{Experiments}
\label{sec:results}

\subsection{Data description}

We conducted experiments on the 2007 NIST Language Recognition Evaluation~(LRE). Our training corpus including  Callfriend datasets, LRE 2003, LRE 2005, SRE 2008 datasets, and development data for LRE07. The total training data is about 37000 utterances. 

The task of interest is closed-set language detection. There are 14 target languages in the testing corpus, which included 7530 utterances with three nominal durations of  30, 10 and 3 seconds.

\subsection{Neural network training}
Audio is converted to 64-dimensional log Mel-filterbank energies with a frame-length of 25 ms, mean-normalized over a sliding window of up to 3 seconds. A frame-level energy-based voice activity detection (VAD) selects features corresponding to speech frames. 

In order to get higher-level abstract representation, we design a deep CNN based on the well-known ResNet-34 architecture \cite{He2016Deep}, as described in Table \ref{resnetconfig}. All the convolutions are with kernel size $k=3\times3$, padding size $p=1$.  Stride size $s=1$ while downsample option in Table \ref{resnetconfig} is set to false,  $s=2$ while true. Followed by the deep CNN, two layer 128-dimensional BLSTM is built to capture the temporal structure, and produces a variable-length sequence of shape $256\times \frac{L}{8}$. After the CNN-BLSTM front-end module, a SAP is designated on top to get the utterance-level representation. 

Therefore, given the input data sequence of shape $64\times L$,  where $L$ denotes variable-length data frames, we finally get 256-dimensional utterance-level representation. The number of output classes is 14.

We use common stochastic gradient descent (SGD) with momentum 0.9 and weight decay 1e-4.  The learning rate is set to 0.1, 0.01,
0.001 and is switched when the training loss plateaus. Since we have no separated validation set, the converged model after the last optimization step is used for evaluation.

In the training stage, the model is trained with a mini-batch size of 128. We design a data loader to generate the variable-length training examples dynamically. For each training step,  an integer $L$ within $\left[ 200 \textrm{,}  1000 \right]$  interval is randomly generated, and each data in the mini-batch is cropped or extended to $L$ frames. Therefore, a dynamic mini-batch of data with the shape of $128 \times 64 \times L$ is generated on-the-fly, and $L$ is a batch-wise variable number. 

In the testing stage, all the 3 seconds, 10 seconds, and 30 seconds duration data is tested on the same model. Because the duration length is arbitrary, we feed the testing speech utterance to the trained neural network one by one.

\subsection{Evaluation}
Table \ref{table:lre07} shows the performance on the 2007 NIST LRE closed-set task.  
The performance is reported in average detection cost $C_{avg}$  and equal error rate (EER).

For our purpose in exploring how to exploit the temporal structure in the neural network and
using BLSTM as part of the front-end module, we first implement two CNN baseline system without any recurrent layer.  

For CNN-TAP system, the front-end CNN is directly connected with a TAP layer to get the utterance-level representation. For CNN-SAP system, the TAP layer is replaced by an SAP layer. Comparing the results of CNN-TAP and CNN-SAP, we can find that the attention-based pooling mechanism can improve the system performance significantly.	

When we apply context-dependent GRU/LSTM as our encoding layer rather than the context-independent TAP or SAP layer, the performance gets much worse. Especially, when the full 30-seconds duration utterance is fed into the CNN-GRU/CNN-LSTM trained within 200 $\scriptsize{\sim}$1000 frames (about 2$\scriptsize{\sim}$10 seconds), the performance drops sharply and almost near random. 

 In previous work \cite{Geng2016End},  ~\textit{Geng et al.} utilized LSTM as the front-end module and introduced an attention-based mechanism to get the utterance-level representation. They only give results on the 3 seconds short duration task, and the performance is not as good as those systems using deep CNN as the front-end module. 
 
In our experiment, first, we introduce BLSTM into the raw CNN-TAP baseline system and implement a tandem CNN-BLSTM TAP system.  It is very interesting that although CNN-BLSTM module introduces much more parameters comparing with the original CNN, it results in slightly degraded performance. It might be the reason that the BLSTM produced output representation are naturally ``unequal''. For example, the last step in the forward direction and the first step in the backward direction may be encoded with more information about the utterance than other time steps.     

Therefore, a simple average operation might not be appropriate, and we replace the TAP layer with the SAP layer. The network architecture of the final tandem CNN-BLSTM SAP system is described in Table \ref{resnetconfig}. The performance is not only much superior to the CNN-BLSTM-TAP system but also better than the CNN-TAP and CNN-SAP baseline system, especially for the 30 seconds long duration task. Moreover, the score-level fusion of system ID2 and ID7 can further improve the performance significantly.

    \section{Conclusions}
    \label{sec:conclusion}
    
In this paper, we implement an attention-based CNN-BLSTM model for utterance-level end-to-end LID. The front-end tandem CNN-BLSTM  module plays a role as local pattern extractor for the variable-length inputs, and the following SAP layer is built on top to get the fixed-dimensional utterance-level representation. The experiment results show the superiority of the tandem CNN-BLSTM SAP system, especially for the long duration task. 

\bibliographystyle{IEEEbib}
\bibliography{refs.bib}

\end{document}